# Ultrafast Stern-Gerlach and Anomalous Bragg Diffraction Regimes of Low-energy Free Electron Interaction with Light

Yongcheng Ding[1], Zirui Zhao[2], Bin Zhang[2,3], Qiaofei Pan[4,5], Mikel Sanz[1], and Yiming Pan[2]

1. Department of Physical Chemistry, University of the Basque Country UPV/EHU, Apartado 644, 48080 Bilbao, Spain
2. State Key Laboratory of Quantum Functional Materials, School of Physical Science and Technology and Center for Transformative Science, ShanghaiTech University, Shanghai 200031, China
3. School of Electrical Engineering - Physical Electronics, Center of Laser-Matter Interaction, Tel Aviv University, Ramat Aviv 69978, Israel
4. Institute of Precision Optical Engineering, School of Physics Science and Engineering, Tongji University, Shanghai, 200092, China
5. MOE Key Laboratory of advanced micro-structure materials, Shanghai, 200092, China

**Abstract**

Recent advances in photon-induced near-field electron microscopy (PINEM) have significantly impacted allied disciplines such as laser-driven accelerators and free electron radiations, collectively fostering the emergence of free-electron quantum optics (FEQO). A central objective of FEQO is to achieve coherent optical control of free electrons, analogous to light manipulation of atoms in atom optics. Motivated by this analogy, we propose an ultrafast Stern-Gerlach (USG) regime for low-energy quantum electron wavepacket (QEW), which crucially incorporates the effects of second-order dispersion inherent to slow electrons. We demonstrate that the USG diffraction induces spectral splitting and shifting of the QEW via a longitudinal electric field gradient, with the two dominant truncated sidebands forming a pseudospin degree of freedom for an effective "two-level" electron. Furthermore, by examining the wave-particle duality of the QEW during light interaction, we identify a dispersion-induced anomalous Bragg diffraction regime. This regime exhibits a distinct spectral pattern, differentiating it from these reported PINEM (Raman-Nath), dielectric laser accelerators (DLA), anomalous PINEM, and Bragg diffraction regimes. Our study provides a comprehensive classification for light-induced diffraction regimes for both swift and slow electrons. These findings underscore the pivotal role of slow-electron dispersion and duality nature in free-electron optics, offering promising avenues for electron wavefunction quantum engineering ultrafast interferometers.

Controlling free electron with light is challenging, but it holds immense potential for advancing quantum technologies. Achieving precise trapping and shaping of electron wavefunctions is essential for establishing free electrons as a feasible platform for quantum simulations, imaging and information processing. Historically, electrons have been central to the development of quantum mechanics. The landmark Stern-Gerlach experiment of 1922[1], employing a beam of neutral silver atoms, provided the first direct experimental evidence of quantum superposition and angular momentum quantization, thereby establishing the concept of spin unexpectedly. Building upon this foundation, the rise of atomic optics in the late twentieth century propelled the field of atomic, molecular, and optical (AMO) physics. Notable progress was the proposal in 1987 [2] and subsequent experimental realization of the optical Stern-Gerlach (OSG) effect [3] in 1992, in which atomic wavepackets were coherently split by the optical field gradients without invoking genuine spin degree of freedom. In atom optics, light-atom diffractions are conventionally categorized into three regimes [4]: the Raman-Nath regime, characterized by transient phase modulation; the Bragg regime, defined by limited momentum-selective diffraction; and the OSG regime, governed by gradient-induced atomic splitting. These regimes underpin a wide range of coherent quantum control techniques in AMO systems.

By analogy with atomic systems, one might expect that free-electron and light interactions give rise to analogous diffraction regimes, albeit exhibiting distinct physical characteristics. Coherent control of free-electron beams is considerably more demanding than for atoms, due to the high-quality electron beam requirements imposed by applications in electron accelerators and free electron lasers. Recent advances in ultrafast electron microscopy and dielectric laser accelerators (DLA) have converged within the emerging field of free-electron quantum optics (FEQO) [5]. Within the FEQO framework, light-electron interactions provide access to regimes that probe the wave-particle duality nature of both light and electron, in close analogy to those found in atom optics. To date, four principal diffraction regimes have been identified: (i) PINEM regime [6–9], characterized by the formation of multiphoton sidebands through coherent phase modulation, analogous to Raman-Nath diffraction. (ii) DLA regime [10–13], where free electrons undergo net acceleration and multiphoton sideband overlap induces classical particle-like beam dynamics [14]. (iii) Bragg diffraction regime [15–17], where second-order dispersion of slow electrons enables momentum-selective light interaction. This regime effectively maps to a two-level Rabi system, with the dynamics transitioning to Raman-Nath behavior governed by the Klein-Cook parameter [16]. The last regime is (iv) anomalous PINEM (APINEM) diffraction[12,18], manifested as a periodically-modulated spectrum for highly pre-chirped electron wavepackets undergoing wave-particle transition during interaction [12].

Despite analogues with atomic systems, free-electron optics exhibits fundamentally distinct features. In light-electron interactions, the particle-like nature of electrons becomes paramount, underpinning key mechanisms in free electron lasers and laser-driven accelerators. When quantum electron wavepackets (QEWs) are substantially smaller than the optical wavelength, the resulting overlap of multiphoton sidebands lead to a shift in the central energy, indicating that the transition from PINEM to DLA regime is a crossover rather than a sharp phase transition. On the other hand, dispersion and chirp in QEWs modify the Bragg regime by truncating multi-level Rabi dynamics into an isolated two-level system [15,17,19], giving rise to the notion of a "free-electron qubit" unique to slow electrons, analogous to an atomic qubit system. However, the optical Stern-Gerlach effect, wherein a gradient field induces spatial splits of atomic wavepackets associated with different pseudospin states, has yet to be realized in free-electron optics. **This notable absence raises a fundamental question: why has such a pivotal diffraction regime in atom optics not manifested with free electrons?** The core challenge lies in synthesizing a pseudospin degree of freedom in response to light while preserving coherence amid dispersion and multiphoton interaction. A promising theoretical pathway emerges from: within the Bragg regime (see Fig. 1), strong dispersion truncates the multiphoton sidebands to an effective two-level system, while the optical field gradient serves as an inhomogeneous pseudomagnetic field. For a slow electron wavepacket with sufficiently low energy, this gradient induces a spectral splitting of the sidebands, thereby establishing a direct route to the regime we termed the ultrafast Stern-Gerlach (USG) regime.

In this Letter, we propose the USG regime, completing the classification of ultrafast light-electron interactions regimes alongside the previously established PINEM, APINEM, DLA, and Bragg diffraction regimes. Our construction exploits the second-order dispersion of low-energy QEWs within the Bragg regime. Interaction with an optical field gradient then induces coherent and controllable spectral splitting of the QEW, accompanied by distinctive split sidebands. Notably, when the QEW's sideband width broadens closer to light quanta, a distinct anomalous Bragg regime emerges: even in the absence of pre-chirping, the modulated sideband spectral distribution exhibits a periodic micro-structure rather than the sideband splitting of USG diffraction. With the discovery of both the USG and anomalous Bragg regimes, we resolve previously unclassified interaction regimes and present a complete taxonomy of light-free-electron diffractions (Table I) beyond these regimes found in atom optics. Our findings underscore the combined roles of wave-particle duality, energy dispersion, and free-space chirping in shaping ultrafast electron wavefunction. By systematically delineating these regimes, our work provides a unified perspective on light-electron interactions, with broad implications for free-electron quantum optics, ultrafast electron spectroscopy and imaging, accelerator physics, and next-generation free electron radiative schemes.

**Light-induced Ultrafast Stern-Gerlach regime.** To describe how light induces a synthetic Stern-Gerlach effect in free electrons, we begin with the minimal coupling Hamiltonian for a nonrelativistic, low-energy electron interacting with a light field: $H = \frac{(p-eA)^2}{2m}$, where $p$ is the momentum and $A$ is the vector potential associated with the incident optical field. In the presence of a longitudinally structured electromagnetic environment, such as that provided by an optical grating, the vector potential and corresponding electric field take the form: $A(z,t) = \frac{E_0(z)}{\omega_L}\cos(\omega_L t - qz)$ and $E(z) = -\partial A/\partial t = E_0(z)\sin(\omega_L t - qz)$, where $\omega_L$ is the laser frequency and q denotes the longitudinal photon wavevector. This travelling-wave configuration co-propagates with the QEW along the grating axis, facilitating efficient coherent longitudinal modulation for the QEW. The time-dependent Schrödinger equation (TDSE) governs the dynamics of the QEW in this optical background. Two primary diffraction regimes emerge, dictated by the interplay between interaction strength and electron dispersion: Raman-Nath regime and Bragg regime. The Raman-Nath regime, also known as PINEM regime, typically realized for high-energy electrons (e.g., 200 keV), where electron dispersion is negligible, exhibiting multiphoton sideband generation via photon exchange with the optical field. Conversely, the Bragg regime is accessed at lower electron energies (e.g., 100 eV), where second-order dispersion dominates, and the weak electron-light interaction is confined into few photons scattering process.

In a limit case, the systems evolve toward an effective two-level Rabi oscillation between two sidebands [15–17], forming the precursor for the USG regime explored in this work.

We first construct a Dirac-like equation to describe light-induced electron dynamics in the Bragg regime, where discrete momentum states corresponding to these photon sidebands form a synthetic momentum lattice. Around a central momentum $k = k_0$, the electron wavevector is further shifted such that $\delta k \to \delta k \pm \frac{q}{2}$, where q denotes the quantum recoil momentum, $\delta k$ is a continuous variable representing the momentum spread of the QEW. The Bragg diffraction condition selects two dominant components - $\left|+\frac{q}{2}\right\rangle$ and $\left|-\frac{q}{2}\right\rangle$, which constitute an effective two-level system. The phase-matching condition $v_g = v_p$, imposes $\frac{\hbar k_0}{m} = \frac{\omega_L}{q}$, aligning the electron group velocity $v_g$ with the optical phase velocity $v_p$, and setting the energy calibration the two-level system in the synthetic lattice. By adiabatically eliminating higher-order sidebands ($\left|\pm\frac{3q}{2}\right\rangle, \cdots$), we derive coupled-mode equations for the "two-level" QEW, which is governed by the following Dirac equation[20]:

$$i\hbar\, \partial_t |\widetilde{\Psi}\rangle = \mathrm{H_D} |\widetilde{\Psi}\rangle, \tag{1}$$

where the spinor wavefunction is $|\widetilde{\Psi}\rangle = c_0 \left|+\frac{q}{2}\right\rangle + c_1 \left|-\frac{q}{2}\right\rangle = [c_0, c_1]^T$, and the effective Dirac Hamiltonian reads:

$$\mathrm{H_D} = \frac{\hbar^2 \delta k^2}{2m} + \frac{\hbar^2 q}{2m} \delta k \sigma_z - \Omega(z,t)\sigma_x. \tag{2}$$

Here, $\sigma_z, \sigma_x$ denote the Pauli matrices acting in the pseudospin space, and the Rabi frequency is defined as $\Omega = eE_0 \hbar\, k_0/2m\omega_L$. The two-level truncation remains valid under the condition $|\Omega| \ll \hbar^2 q^2/8m$, ensuring isolation of the two sidebands. Spatiotemporal variation in $\Omega(z,t)$ may be introduced either via the envelope or phase profile of the incoming optical field. For instance, a spatially dependent phase $\theta(z,t)$ of the light field modifies the coupling term as $\Omega \exp(-i\theta\sigma_z)\sigma_x$, enabling topological phenomena such as the Jackiw-Rebbi bound state [20]. In the present work, we focus on the realization of an inhomogeneous Rabi amplitude induced by a longitudinal field gradient. This field profile acts as a pseudomagnetic field gradient (the inset of Fig. 1a), leading to sideband splitting of the QEW, constituting a light-induced Stern-Gerlach diffraction in momentum (energy) domain.

Figure 1a depicts the proposed experimental setup. A QEW is generated from a photoemission cathode and shaped by a femtosecond laser pulse on a grating. As the QEW passes through a laser-illuminated optical grating (like a periodically structured travelling-wave conical tube), it encounters a spatially varying longitudinal electric field, thereby realizing the minimal coupling Hamiltonian for the resonant interaction in a structured optical background. In Bragg diffraction regime, the weak light-electron interaction gives rise to an effective two-level system (Fig. 1b). Consequently, the QEW acquires pseudospin dynamics analogous to those of an electron in an inhomogeneous magnetic field, giving rise to the USG diffraction. When the QEW sideband state is initial prepared as a coherent superposition of $\left|\pm\frac{q}{2}\right\rangle$, the eigenstate of $\sigma_x$, the two pseudospin components experience equal and opposite optical gradient forces. This induces a momentum-space sideband splitting of the QEW (Fig. 1c), which can be resolved using high-resolution electron energy spectrometers.

**Splitting of QEW via optical gradient forces.** We investigate the quantum dynamics of a free-electron wavepacket by numerically solving the TDSE with the minimal coupling Hamiltonian under phase matching and weak field condition. Phase matching with the optical field is achieved via a grating, which modulates the longitudinal electric field experienced by the electron. Under a spatially uniform field profile, corresponding to a constant Rabi frequency $\Omega$ (Eq. 2), the QEW exhibits coherent oscillation between the momentum sideband states $|+q/2\rangle$ and $|-q/2\rangle$. To initialize the two-level dynamics, a Gaussian QEW is prepared, occupying only at $|+q/2\rangle$, with a central energy of 100 eV (group velocity $v_g = 0.02c$) and momentum width $\Delta_k = 0.02q$. The interacting optical field has a photon energy $\hbar\omega_L = 6.2$ eV (wavelength $\lambda = 200$ nm), and is phase matched a well-nanofabricated grating of 4 nm period. The field amplitude is set to $E_0 = 10^8$ V/m, and the interaction duration is $T = 25$ ps, corresponding to approximately eight Rabi cycles. As shown in Fig. 2a, the simulated momentum distributions reveal oscillatory population transfer between the two dominant sidebands. Nevertheless, residual population leakage into higher-order sidebands is evident. These deviations from the ideal two-level dynamics are attributed to the imperfect trap offered by the slow-electron dispersion compared with the intensity of the driving field, where the Klein-Cook parameter $Q \simeq 1.4$ in our simulation. The high-order sidebands would be negligible when $Q \gg 1$.

To induce momentum splitting of the QEW via an optical gradient force, we apply a spatially varying longitudinal electric field with a central amplitude $E_0(z = 0) = 10^8$ V/m, increasing linearly from 0 V/m at $z = -100$ nm to $2 \times 10^8$ V/m at $z = 100$ nm. This field profile introduces an optical gradient force acting on the pseudospin degrees of freedom, leading to momentum deflection. To characterize the dynamics in the combined $(\delta k, z)$ phase space, we employ the Wigner function distribution (WFD): $W(z,p) = \frac{1}{\pi\hbar}\int \Psi_p^*(p+q)\Psi_p(p-q)e^{-\frac{2izq}{\hbar}}dq$, which allows direct visualization of the QEW evolution. The QEW is initialized in an eigenstate of $\sigma_x$: $|+\rangle = (1/\sqrt{2})(|+q/2\rangle + |-q/2\rangle)$, such that the two pseudospin components experience opposite forces in the presence of the optical gradient. As shown in Fig. 2b, the resulting WFD exhibits a spectral shift in momentum space, characteristic of gradient-induced acceleration and consistent with the PINEM-to-DLA crossover behavior for swift electron wavepackets [12,13]. The observed shift is well described by the effective Hamiltonian containing only the spatially varying coupling term $\Omega(z)\sigma_x$. Additionally, population leakage into the higher-order sidebands such as $|\pm 3q/2\rangle$ remains negligible, confirming the validity of the Bragg diffraction regime in the presence of spatial inhomogeneity.

Figure 2c illustrates the manifestation of the USG effect for free electrons in the Bragg regime, where both the oscillatory sidebands (Fig. 2a) and spectral shift (Fig. 2b) are present simultaneously. The QEW is initialized in the state $|0\rangle = |+q/2\rangle$, and the TDSE simulations reveal simultaneous sideband population transfer and splitting of each eigencomponent. The spatial gradient in the light-matter coupling, $\nabla\Omega(z,t)$, generates a net momentum shift given by $\int_0^T \nabla\Omega(z,t)dt$, yielding a peak separation of $\Delta k = 16\pi\ \mu m^{-1}$. The resulting sub-wavepacket trajectories align well with the predicted USG deflection. Analysis of the Wigner function distribution further confirms the presence of four well-separated spectral lobes centered at $\delta k = \pm \frac{q}{2} \pm \frac{1}{2} \int_0^T \nabla\Omega(z,t)dt$, oriented nearly parallel to the z-axis. While the ideal model predicts a linear dependence on the optical field gradient, this behavior is only asymptotically recovered in the weak-gradient limit. Under strong-field coupling condition, deviations arise due to population leakage into off-resonant states and spectral distortion of the QEW (see Fig. 2d).

Beyond validating the USG regime (Fig. 2e), we evaluate our setup's performance as a free-electron quantum simulator of Dirac physics. By filtering near $\delta k = \pm q/2$, we compare energy-resolved electron spectroscopy derived from TDSE simulations to those constructed from the Dirac-type Hamiltonian. The close agreement between two approaches confirms the simulator's fidelity. Notably, simulation parameters were deliberately selected to expose imperfections and test the platform's robustness under experimentally realistic conditions. These results suggest that with suitable parameter optimization, the system can serve as a versatile simulator for relativistic pseudospin dynamics.

**Anomalous Bragg diffraction transitioning from USG regime.** In conventional optical SG experiments, the wavepacket must be tightly confined within one period of the optical potential, necessitating tightly spatial confinement to atomic beams [4]. This raises a fundamental question regarding the degree of localization required for a QEW to exhibit ultrafast Stern-Gerlach dynamics. In the simulation shown in Fig. 2c, the QEW has a spatial width of approximately 80 nm, yet the simulation reveals clear momentum-space splitting. This observation indicates that spatial confinement is not prerequisite in free electron systems. Instead, the key requirement is sufficiently small width of the two sidebands to preserve orthogonality between the eigenstates $|\pm q/2\rangle$. Specifically, a lower bound of $\Delta_k \geq q/4$ ensures adequate pseudospin separation by suppressing interference between the sidebands. This stands in stark contrast to atom optics, where spatial confinement of the atom beams is crucial for observing the optical SG effect [4].

To further explore the consequence of this spatial confinement, we consider a more tightly localized QEW, with a width of $\Delta_k = 0.15q$, wider than the QEW in Fig. 2c, while the eigenstates $|\pm q/2\rangle$ still remain separative. Contrary to the expected USG splitting, the wavepacket undergoes distorted Rabi-like oscillations that are symmetric about the central axis (Fig. 3a). This behavior is counterintuitive and nontrivial, considering that increasing the momentum width $\sigma_k$ is generally presumed to reinforce USG behavior rather than cancelling it. Examination of the Wigner function distribution reveals nonclassical phase-space patterns, including oscillatory fringes and lobe symmetry, analogous to characteristic of anomalous PINEM [12] that is realized in DLA regime with highly pre-chirped QEWs. In fact, this anomalous diffraction arises due to the fact that slow electrons, on the order of 100 eV, with narrower spatial width are much more sensitive to second-order dispersion of the QEW during the interaction. Consequently, the transition from the USG regime to an anomalous Bragg regime occurs via a smooth crossover, without a sharp threshold. As shown in Fig. 3b, the evolution of the momentum distribution illustrates this crossover. Our results suggest that when such anomalous Bragg patterns are observed experimentally, further reduction of the sideband width, i.e., increasing spatial coherence of QEW, may be necessary to fully access the USG regime.

Notably, we address that this anomalous Bragg regime reported here is distinct from APINEM regime, despite superficial similarities in their spectral structures. The key distinction lies in initial conditions of the QEW. In APINEM [12,21], the free electron pulse must be highly pre-chirped to enable multiphoton sideband resolution in phase space, as characterized by the Wigner function distribution. In contrast, the QEW in our configuration is position-localized but momentum-delocalized, similar to particle-like initial states typical of DLA scenarios. The spread momentum wavepacket feels distinct energy gap $E_{\pm}(\delta k) = \pm\sqrt{|\Omega|^2 + \hbar^4 q^2 \delta k^2/4m^2}$ depicted by the Dirac equation Eq. 2 in the momentum domain, exhibiting distinct Rabi period $T(\delta k) = 2\pi/E_{\pm}(\delta k)$ for the two sidebands forming the synthetic two-level space. Despite this, it exhibits exotic diffraction behavior. Unlike the USG patterns, the resulting diffraction patterns show negligible dependence on the presence or absence of an optical gradient. These features motivate our classification of this regime as anomalous Bragg diffraction, a previously unrecognized regime of free-electron-light interaction. Table I provides a comparative summary of all diffraction regimes identified in FEQO, highlighting their operational conditions and distinctions from their atom optics analogues.

Looking ahead, one compelling extension to investigating free-electron diffraction under illumination by quantum light sources, including nonclassical Fock states, squeezed coherent states, and entangled photons. Such studies would move beyond the semiclassical framework, enabling quantum simulation of coupled two-level QEW and its spinor dynamics under quantized fields. Furthermore, by employing post-selection of the QEW [22], it may be possible to implement coherent quantum control over nonclassical light states, with the protocols of weak measurements [14]. Finally, the experimental realization of the USG regime offers a promising route for constructing free-electron interferometers and quantum sensors [16,17,23–32], quantum-coherent platforms for precision sensing and ultrafast imaging with momentum-resolved capabilities.

**Conclusion.** In short, we have proposed and analyzed an ultrafast Stern-Gerlach (USG) effect for a low-energy quantum electron wavepacket (QEW). By formulating an effective Dirac equation that maps the interaction onto a two-level system, we demonstrated that a spatially varying light-matter coupling can induce spinor-dependent splitting of the QEW. Numerical simulations confirm this USG behavior, showing clear momentum-space deflection in excellent agreement with theoretical predictions. We also identified a previously unexplored diffraction regime, termed the anomalous Bragg regime, that exhibits rich, nonclassical spectral patterns. These findings emphasize the phase-space structure of QEW, energy dispersion, and wave-particle coherence, often negligible in AMO settings, play a decisive role in shaping light-free-electron interactions. Taken together, our findings not only complete the classification of light-induced diffraction regimes but also lay a foundation for future developments in ultrafast quantum sensing, coherent control, and quantum simulation using free electrons.

**Acknowledgement**

Y. D. thanks the European Commission for a Marie Curie PF grant (No. 101204580 FELQO). Y. P. acknowledges the support of the NSFC (No. 2023X0201-417-03) and the fund of the ShanghaiTech University (Start-up funding).

**References**

1.  Gerlach, W. & Stern, O. Der experimentelle Nachweis der Richtungsquantelung im Magnetfeld. *Zeitschrift für Physik* **9**, 349–352 (1922).
2.  Cook, R. J. Optical Stern-Gerlach effect. *Phys Rev A* **35**, 3844–3848 (1987).
3.  Sleator, T., Pfau, T., Balykin, V., Carnal, O. & Mlynek, J. Experimental demonstration of the optical Stern-Gerlach effect. *Phys Rev Lett* **68**, 1996–1999 (1992).
4.  Meystre, P. *Quantum Optics*. (Springer International Publishing, Cham, 2021). doi:10.1007/978-3-030-76183-7.
5.  Ruimy, R., Karnieli, A. & Kaminer, I. Free-electron quantum optics. *Nat Phys* **21**, 193–200 (2025).
6.  Barwick, B., Flannigan, D. J. & Zewail, A. H. Photon-induced near-field electron microscopy. *Nature* **462**, 902–906 (2009).
7.  Park, S. T., Lin, M. & Zewail, A. H. Photon-induced near-field electron microscopy (PINEM): theoretical and experimental. *New J Phys* **12**, 123028 (2010).
8.  Garcia De Abajo, F. J., Asenjo-Garcia, A. & Kociak, M. Multiphoton absorption and emission by interaction of swift electrons with evanescent light fields. *Nano Lett* **10**, 1859–1863 (2010).
9.  Feist, A. *et al.* Quantum coherent optical phase modulation in an ultrafast transmission electron microscope. *Nature* **521**, 200–203 (2015).
10. Breuer, J. & Hommelhoff, P. Laser-Based Acceleration of Nonrelativistic Electrons at a Dielectric Structure. *Phys Rev Lett* **111**, 134803 (2013).
11. Peralta, E. A. *et al.* Demonstration of electron acceleration in a laser-driven dielectric microstructure. *Nature* **503**, 91–94 (2013).
12. Pan, Y., Zhang, B. & Gover, A. Anomalous Photon-Induced Near-Field Electron Microscopy. *Phys Rev Lett* **122**, (2019).
13. Gover, A. & Pan, Y. Dimension-dependent stimulated radiative interaction of a single electron quantum wavepacket. *Phys Lett A* **382**, 1550–1555 (2018).
14. Pan, Y. *et al.* Weak measurements and quantum-to-classical transitions in free electron–photon interactions. *Light Sci Appl* **12**, 267 (2023).
15. Eldar, M., Chen, Z., Pan, Y. & Krüger, M. Self-Trapping of Slow Electrons in the Energy Domain. *Phys Rev Lett* **132**, 035001 (2024).
16. Pan, Y., Zhang, B. & Podolsky, D. Low-energy Free-electron Rabi oscillation and its applications. (2023).
17. Karnieli, A. & Fan, S. Jaynes-Cummings interaction between low-energy free electrons and cavity photons. *Sci Adv* **9**, eadh2425 (2023).
18. Zhou, J., Kaminer, I. & Pan, Y. *Quantum Emergence of Linear Particle Accelerator and Anomalous Photon-Induced Near-Field Electron Microscopy in a Strong Coupling Regime*.
19. Pan, Y., Zhang, B. & Podolsky, D. Low-energy Free-electron Rabi oscillation and its applications. *arXiv preprint arXiv:2304.12174* (2023).
20. Pan, Y. *et al.* Free electron topological bound state induced by light beam with a twisted wavefront. (2024).
21. Lin, K. *et al.* Ultrafast Kapitza-Dirac effect. *Science (1979)* **383**, 1467–1470 (2024).
22. Aharonov, Y., Albert, D. Z. & Vaidman, L. How the result of a measurement of a component of the spin

of a spin- $1/2$ particle can turn out to be 100. *Phys Rev Lett* **60**, 1351–1354 (1988).
23. Bucher, T. *et al.* Free-electron Ramsey-type interferometry for enhanced amplitude and phase imaging of nearfields. *Sci Adv* **9**, (2023).
24. Ruimy, R., Gorlach, A., Mechel, C., Rivera, N. & Kaminer, I. Toward Atomic-Resolution Quantum Measurements with Coherently Shaped Free Electrons. *Phys Rev Lett* **126**, 233403 (2021).
25. Gorlach, A. *et al.* Ultrafast non-destructive measurement of the quantum state of light with free electrons. in *Conference on Lasers and Electro-Optics* FF2I.4 (Optica Publishing Group, Washington, D.C., 2021). doi:10.1364/CLEO_QELS.2021.FF2I.4.
26. Gaida, J. H. *et al.* Attosecond electron microscopy by free-electron homodyne detection. *Nat Photonics* **18**, 509–515 (2024).
27. Bucher, T. *et al.* Coherently amplified ultrafast imaging using a free-electron interferometer. *Nat Photonics* **18**, 809–815 (2024).
28. Feist, A., Yalunin, S. V., Schafer, S. & Ropers, C. High-purity free-electron momentum states prepared by three-dimensional optical phase modulation. *Phys Rev Res* **2**, (2020).
29. Echternkamp, K. E., Feist, A., Schäfer, S. & Ropers, C. Ramsey-type phase control of free-electron beams. *Nat Phys* **12**, 1000–1004 (2016).
30. Priebe, K. E. *et al.* Attosecond electron pulse trains and quantum state reconstruction in ultrafast transmission electron microscopy. *Nat Photonics* **11**, 793–797 (2017).
31. Karnieli, A. *et al.* Quantum sensing of strongly coupled light-matter systems using free electrons. *Sci Adv* **9**, (2023).
32. Asban, S. & García de Abajo, F. J. Generation, characterization, and manipulation of quantum correlations in electron beams. *npj Quantum Inf* **7**, 42 (2021).

Figures:

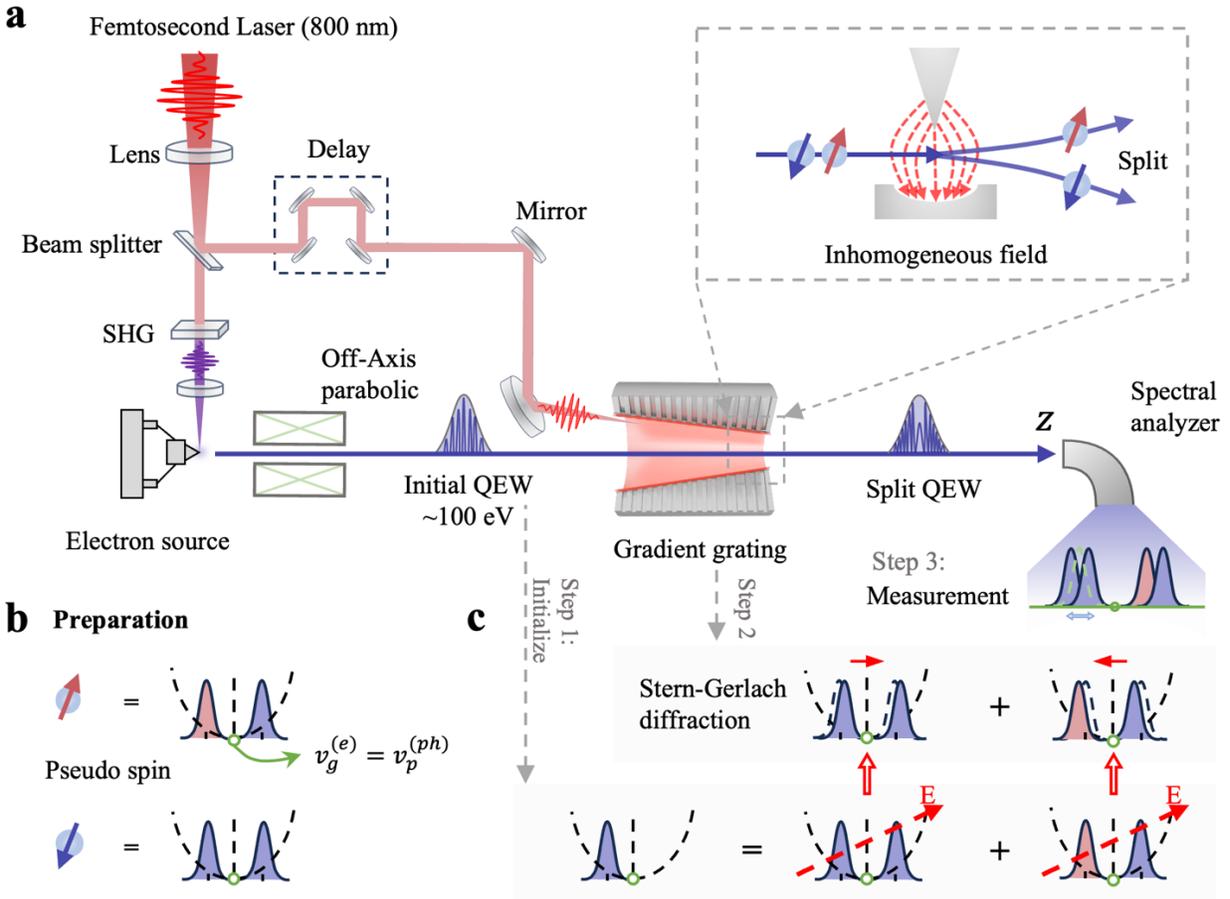

**Figure 1.** Setup and realization of Ultrafast Stern-Gerlach (USG) diffraction for slow electrons. (a) Schematic of the experimental setup as compared with ultrafast electron microscope, featuring a photoelectron gun, a laser-illuminated grating, and an electron energy spectroscopy system. (b) Quantum electron wavepacket (QEW) state preparation. Under phase-matching conditions, the QEW is assumed to be prepared in a superpositions of $|\pm\rangle$, the eigenstates of the $\sigma_x$ operator, leading to opposite spectral splitting. (c) Illustration of USG splitting. A QEW initially prepared in $|0\rangle$, the 50:50 superposition of $|\pm\rangle$, undergoes momentum separation, with spectral shear left or right upon interaction with the grating's gradient field. The resulting diffraction pattern is measured using electron spectral analyzer.

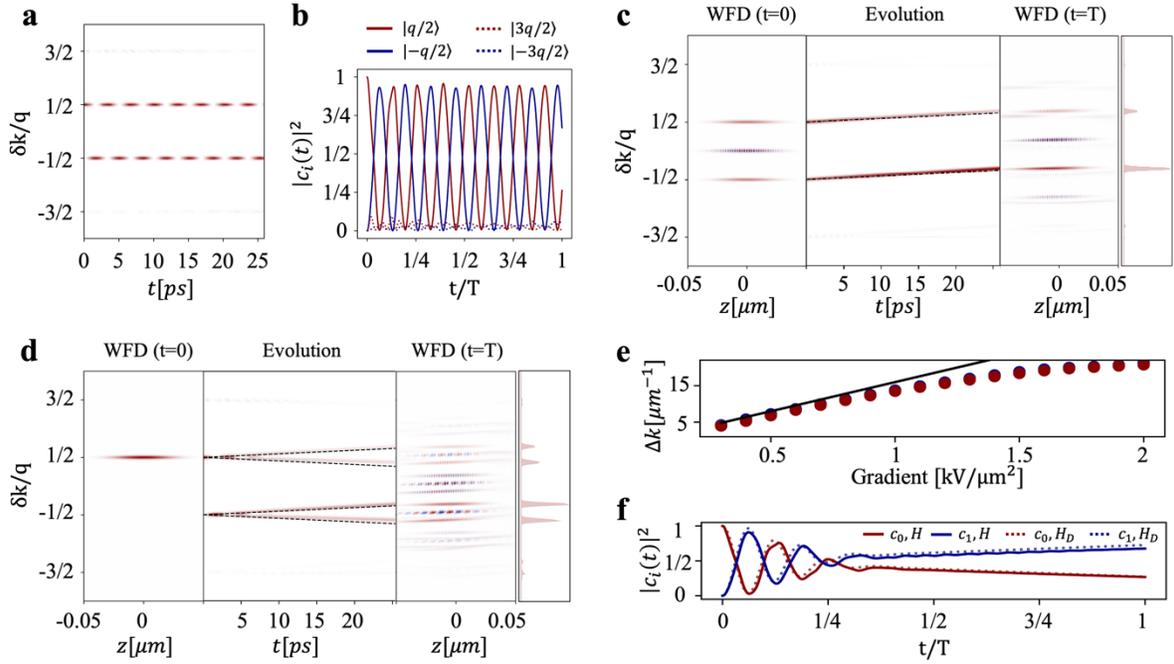

**Figure 2.** TDSE simulation of the slow QEW evolution in the Bragg regime. (a) QEW evolution in momentum space under a spatially uniform light field. (b) Rabi oscillation in the effective two-level system, with populations reconstructed by simulated spectra. Leakage to higher sidebands is negligible. (c) QEW evolution under a light field with a spatial gradient. As a calibration, the QEW is initialized in $|+\rangle$, showing a net spectral shift after interaction. Black dashed lines indicate the ideal sideband evolution. (d) Stern-Gerlach diffraction with a gradient field, initialized with $|0\rangle = (1/\sqrt{2})|\pm\rangle$. (e) Spectral split between the $|0\rangle$ (red) and $|1\rangle$ (blue) sidebands, in consistent with the ideal split (black solid line) under varying gradient strengths. (f) Sideband populations on $|0\rangle$ and $|1\rangle$ are simulated using the TDSE (solid) and the effective Dirac equation (dotted).

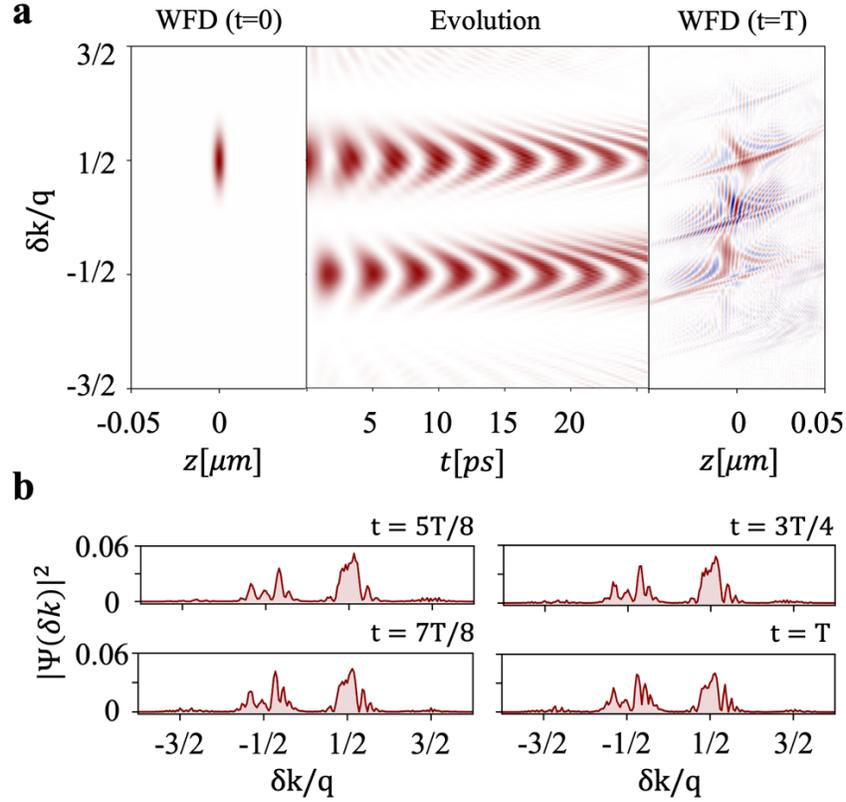

**Figure 3.** TDSE simulation of anomalous Bragg diffraction for slow QEWs with broad momentum sideband width. (a) The diffraction pattern resembles Bragg diffraction but features a chirping-dependent momentum distribution for each sideband, forming a characteristic "fishbone" structure. (b) QEW snapshots taken at time interval of $T/8$ show near-identical profiles, indicating approximate periodicity of the dynamics despite the presence of pre-chirping and field gradient.

| Relative velocity / Wave-particle duality | Fast electron $\beta \sim 1$ | Slow electron $\beta \ll 1$ |
|---|---|---|
| Plane wave limit $\Delta_z \gg \beta\lambda$ | PINEM/ Raman-Nath *multi-level Rabi oscillation* | Bragg *Rabi oscillation* Ultrafast Stern-Gerlach *momentum splitting* Anomalous Bragg |
| Highly chirped QEW $\Delta_z \ll \beta\lambda, \Delta_z(t) > \beta\lambda$ | APINEM *periodic spectral bunching* | APINEM *slow-electron dispersion* |
| Point-particle limit $\Delta_z \ll \beta\lambda$ | DLA *classical acceleration* | N/A |

**TABLE 1.** Classification of diffraction regimes for slow and fast quantum electron wavepackets (QEWs) interacting with light. The regimes are distinguished by the QEW size, given by the intrinsic size $\Delta_z = 1/2\Delta_k$, or by chirped size $\Delta_z(t)$ [12], and relative velocity ($\beta \ll 1$ for slow electrons, $\beta \sim 1$ for fast electrons). For fast QEWs, diffraction is categorized into PINEM, DLA, and APINEM regimes. In the plane-wave limit, regimes include the Raman-Nath (also known as PINEM, showing multi-level Rabi oscillation or quantum random walk) and Bragg (showing Rabi oscillation) and Ultrafast Stern-Gerlach regimes (showing momentum splitting), analogous to classifications in atom optics. For slow QEWs, Bragg diffraction becomes richer. For near-plane-wave QEWs, diffraction patterns exhibit features such as anomalous Bragg diffraction (showing fishbone-like momentum distribution for each sideband) due to slow-electron dispersion and chirping. Acceleration is generally not applicable (N/A) for slow QEWs. When the wavepacket is highly chirped: $\Delta_z(t) > \beta\lambda, \Delta_z \ll \beta\lambda$, APINEM also appears for slow electrons. This work elucidates the Bragg regime and introduces ultrafast SG and anomalous Bragg diffraction.

# Supplementary Material:
# Ultrafast Stern-Gerlach and Anomalous Bragg Diffraction Regimes of Low-energy Free Electron Interaction with Light

Yongcheng Ding[1], Zirui Zhao[2], Bin Zhang[2,3], Qiaofei Pan[4,5], Mikel Sanz[1], and Yiming Pan[2]

1. Department of Physical Chemistry, University of the Basque Country UPV/EHU, Apartado 644, 48080 Bilbao, Spain
2. State Key Laboratory of Quantum Functional Materials, School of Physical Science and Technology and Center for Transformative Science, ShanghaiTech University, Shanghai 200031, China
3. School of Electrical Engineering - Physical Electronics, Center of Laser-Matter Interaction, Tel Aviv University, Ramat Aviv 69978, Israel
4. Institute of Precision Optical Engineering, School of Physics Science and Engineering, Tongji University, Shanghai, 200092, China
5. MOE Key Laboratory of advanced micro-structure materials, Shanghai, 200092, China

This pdf file includes three sections and four supplementary figures.

# 1. Derivation of Dirac Equation for low-energy "two-level" quantum electron wavepackets

Starting from the minimal Hamiltonian

$$H = \frac{(p - eA)^2}{2m}, \tag{S1}$$

we substitute the vector potential $A = \frac{E_0}{\omega_L}\cos[\omega_L t - qz - \theta(z,t)]$, where the phase parameter $\theta(z,t)$ is supposed to vary slowly, and neglect the ponderomotive potential $e^2 A^2/2m$, we have

$$H = \frac{p^2}{2m} - \frac{eE_0}{m\omega_L}\cos[\omega_L t - qz - \theta(z,t)]p - \frac{eE_0 \hbar q}{2im\omega_L}\sin[\omega_L t - qz - \theta(z,t)]. \tag{S2}$$

With the co-moving unitary operator $U(t) = \exp(-ip\omega_L t/\hbar q)$, we derive the time-independent Hamiltonian

$$\widetilde{H} = \frac{p^2}{2m} - \frac{\omega_L}{q}p - \frac{eE_0}{m\omega_L}\cos(qz + \theta)p + \frac{eE_0 \hbar q}{2im\omega_L}\sin(qz + \theta), \tag{S3}$$

following the transformation $\widetilde{H} = U^\dagger H U - i\hbar U^\dagger \partial_t U$. Correspondingly, the wavefunction becomes $\widetilde{\Psi} = U\Psi$. By using Euler's formula $\cos(qz + \theta) = [e^{i(qz+\theta)} + e^{-i(qz+\theta)}]/2$ and $\sin(qz + \theta) = [e^{i(qz+\theta)} - e^{-i(qz+\theta)}]/2i$, we find

$$\widetilde{H}|k\rangle = \left[\frac{\hbar^2 k^2}{2m} - \frac{\omega_L}{q}\hbar k\right]|k\rangle - e^{i\theta}\left(\frac{eE_0 \hbar k}{2m\omega_L} + \frac{eE_0 \hbar q}{4m\omega_L}\right)|k - q\rangle - e^{-i\theta}\left(\frac{eE_0 \hbar k}{2m\omega_L} - \frac{eE_0 \hbar q}{4m\omega_L}\right)|k + q\rangle, \tag{S4}$$

in the momentum representation with $z = i\hbar\,\partial_p$. By letting $|k| \gg q$, we neglect the detuning term in each coupling, resulting in the effective coupled mode equation

$$i\hbar\,\partial_t|k\rangle = \left(\frac{\hbar^2 k^2}{2m} - \frac{\hbar\omega_L}{q}k\right)|k\rangle - \Omega|k - q\rangle - \Omega^*|k + q\rangle, \tag{S5}$$

where $\Omega = \frac{eE_0 \hbar k}{2m\omega_L}e^{i\theta}$ is proportional to the amplitude of the light field. Since the momentum $k$ is continuous, we set $k = k_0 + \delta k$ and expand the on-site term of Eq. S5 s then we have

$$\begin{aligned}\frac{\hbar^2 k^2}{2m} - \frac{\hbar\omega_L}{q}k &= \frac{\hbar^2(k_0 + \delta k)^2}{2m} - \frac{\hbar\omega_L}{q}(k_0 + \delta k) \\ &= \left[\frac{\hbar^2 k_0^2}{2m} - \frac{\hbar\omega_L}{q}k_0\right] + \left(\frac{k_0}{m} - \frac{\omega_L}{q}\right)\hbar\delta k + \frac{\hbar^2 \delta k^2}{2m}\end{aligned} \tag{S6}$$

With the phase matching condition $\frac{k_0}{m} - \frac{\omega_L}{q} = 0$, we have

$$i\hbar\,\partial_t|\delta k\rangle = \left(-\frac{\hbar^2 k_0^2}{2m} + \frac{\hbar^2 \delta k^2}{2m}\right)|\delta k\rangle - \Omega|\delta k - q\rangle - \Omega^*|\delta k + q\rangle. \tag{S7}$$

By shifting $\delta k \to \delta k \pm q/2$ and cancelling all constants, we have the coupled-mode equation for all sidebands and Dirac equation for two sidebands in the main text.

## 2. TDSE simulation algorithm

To ensure the simulation is also compatible with fast electron, we solve the TDSE

$$i\hbar\, \partial_t \Psi(z,t) = H\Psi(z,t) = (H_0 + H_I)\Psi(z,t), \tag{S8}$$

where the kinetic energy

$$H_0 = \mathcal{E}_0 + v_0(p - p_0) + \frac{(p - p_0)^2}{2\gamma^3 m}, \tag{S9}$$

is corrected by special relativity. In the main text, the initial kinetic energy $\mathcal{E}_0 = 100$ eV, corresponding to the electron velocity $v_0 = \beta c$ with $\beta = 0.02$ and Lorentz factor $\gamma = 1.001$. We have the nanograting with longitudinal vector potential $A(z,t) = -\frac{E_0}{\omega_L}\sin[\omega_L t - qz + \theta(z,t)]$ with amplitude $E_0$, laser frequency $\omega_L$, wavevector $q = 2\pi/\Lambda$ for a grating with a period of $\Lambda$, and the phase $\theta$. Note that this vector potential differs from the form used in the main text and in the first section of the Supplemental Material. Accordingly, we write down the interaction Hamiltonian

$$H_I = -\frac{e}{2\gamma m}[p \cdot A + A \cdot p] \approx -\frac{eA_0}{\gamma m}\sin[\omega_L t - qz + \theta(z,t)] \cdot p, \tag{S10}$$

approximated by $p_0 \gg \hbar q$, where $A_0 = E_0/\omega_L$. We express the QEW as a moving part with an initial phase

$$\Psi(z,t) = \chi(z,t) e^{-\frac{i(\mathcal{E}_0 t - p_0 z)}{\hbar}}.$$

Therefore, LHS of Eq. S8 is

$$i\hbar\, \partial_t \Psi(z,t) = e^{-\frac{i(\mathcal{E}_0 t - p_0 z)}{\hbar}}(\mathcal{E}_0 + i\hbar\, \partial_t)\chi(z,t). \tag{S11}$$

Meanwhile, RHS is approximated to

$$(H_0 + H_I)\Psi(z,t)$$
$$\approx e^{-\frac{i(\mathcal{E}_0 t - p_0 z)}{\hbar}}\left\{\mathcal{E}_0 + v_0 p + \frac{p^2}{2\gamma^3 m} - \frac{eA_0}{\gamma m}\sin[\omega_L t - qz + \theta(z,t)] \cdot (p + p_0)\right\}\chi(z,t).$$

Thus, we simplify the TDSE as

$$i\hbar\,\partial_t\chi(z,t) = \{\frac{p^2}{2\gamma^3 m} + [v_0 - \frac{eA_0}{\gamma m}\sin(\omega_L t - qz + \theta)]p\}\chi(z,t) - \frac{eA_0 p_0}{\gamma m}\sin(\omega_L - qz + \theta)\chi(z,t).$$

Substitute $p = -i\hbar\,\partial_z$, $p_0 = \gamma\beta cm$, and $A_0 = E_0/\omega_L$ yields

$$i\hbar\,\partial_t\chi(z,t) = \{-i\hbar[v_0 - \frac{eE_0}{\gamma m\omega_L}\sin(\omega_L t - qz + \theta)]\,\partial_z\}\chi(z,t) - [\frac{\hbar^2}{2\gamma^3 m}\partial_z^2 + \frac{eE_0\beta c}{\omega_L}\sin(\omega_L t - qz + \theta)]\chi(z,t).$$

This partial differential equation can be simplified, saving memory for numerical simulation as well, if we employ the co-moving frame: $\xi = z - v_0 t$, $t' = t$. Accordingly, we have $\partial_z = \partial_\xi$ and $\partial_t = \partial_{t'} - v_0\,\partial_\xi$. Thus, TDSE becomes

$$i\hbar(\partial_{t'} - v_0\,\partial_\xi)\chi(\xi,\tau_c)$$
$$= (-i\hbar\{v_0 + \frac{eE_0}{\gamma m\omega_L}\sin[q\xi - \theta(\xi)]\}\partial_\xi)\chi(\xi,\tau_c) - \{\frac{\hbar^2}{2\gamma^3 m}\partial_\xi^2 - \frac{eE_0\beta c}{\omega_L}\sin[q\xi - \theta(\xi)]\}\chi(\xi,\tau_c),$$

if $\tau_c = ct'$, by dividing each side by $\hbar c$, we have the simplified equation

$$i\,\partial_{\tau_c}\chi(\xi,\tau_c) = -i\{\beta - \alpha_1\sin[q\xi - \theta(\xi)]\}\partial_\xi\chi(\xi,\tau_c) - \{\alpha_2\,\partial_\xi^2 + \alpha_0\sin[q\xi - \theta(\xi)]\}\chi(\xi,\tau_c),$$

where

$$\alpha_0 = \frac{eE_0\beta}{\hbar\omega_L},\quad \alpha_1 = \frac{eE_0}{\gamma mc\omega_L},\quad \alpha_2 = \frac{\hbar}{2\gamma^3 mc}. \tag{S12}$$

For numerical simulation, we discretize the spatial parameter $\xi$ by uniformly dividing the spatial domain into $N_\xi$ parts with the interval $\delta\xi = (\xi_{\max} - \xi_{\min})/N_\xi$. Therefore, the PDE is reformulated to a difference equation

$$\begin{aligned}
i\,\partial_{\tau_c}\chi(\xi,\tau_c) &= \left[-\frac{\alpha_2}{\delta\xi^2} - \frac{i\beta}{2\delta\xi} + \frac{\alpha_1}{2\delta\xi}e^{i[q\xi+\theta(\xi)]}\right]\chi(\xi+\delta\xi,\tau_c) \\
&+ \left[-\frac{\alpha_2}{\delta\xi^2} + \frac{i\beta}{2\delta\xi} + \frac{\alpha_1}{2\delta\xi}e^{-i[q\xi+\theta(\xi)]}\right]\chi(\xi-\delta\xi,\tau_c) \\
&+ \left\{\frac{2\alpha_2}{\delta\xi^2} - \frac{\alpha_1}{\delta\xi}\cos[q\xi+\theta(\xi)] - \alpha_0\sin[q\xi+\theta(\xi)]\right\}\chi(\ ,\tau_c).
\end{aligned} \tag{S13}$$

To simulate the time evolution, we define the vectorized wavefunction $v(\tau) = [\chi(\xi_1,\tau), \chi(\xi_2,\tau), \ldots, \chi(\xi_{N_\xi},\tau)]^T$, and evolve it by one time step $\delta\tau = (\tau - \tau_0)/N_\tau$ with the Crank-Nicholson integrator

$$\begin{aligned}
v(\tau + \delta\tau) &= U(\tau + \delta\tau, \tau)v(\tau), \\
U(\tau + \delta\tau, \tau) &= [1 + i\delta\tau H(\tau)/2]^{-1}[1 - i\delta\tau H(\tau)/2].
\end{aligned} \quad (S14)$$

In this way, we have the numerical simulation of TDSE. To align with the setup in main text and first section, we set the phase parameter $\theta = \pi/2$, which mimics the quantum dynamics of the effective Dirac Hamiltonian with slow electron.

### 3. Diffraction Regimes and Patterns

Here we present numerical simulations of the minimal coupling Hamiltonian under various QEW settings, illustrating different diffraction regimes and patterns, see Figs. S1-S4. In the main text, we refer to the regime characterized by a slow electron and a broadened momentum sideband as the **Anomalous Bragg regime**, regardless of whether a light-field gradient is present. In Fig. S1, we plot the QEW profiles using the light fields from Fig. 2a (without gradient) and Fig. 3 (with gradient) in the main text, with the momentum sideband width set to $\Delta_k = 0.15q$. The profiles at each time step nearly overlap and are barely distinguishable, indicating that dispersion dominates. This distorts the central momentum and gives rise to a fishbone-like Rabi oscillation, rather than a clearly resolved splitting, even in the presence of a gradient.

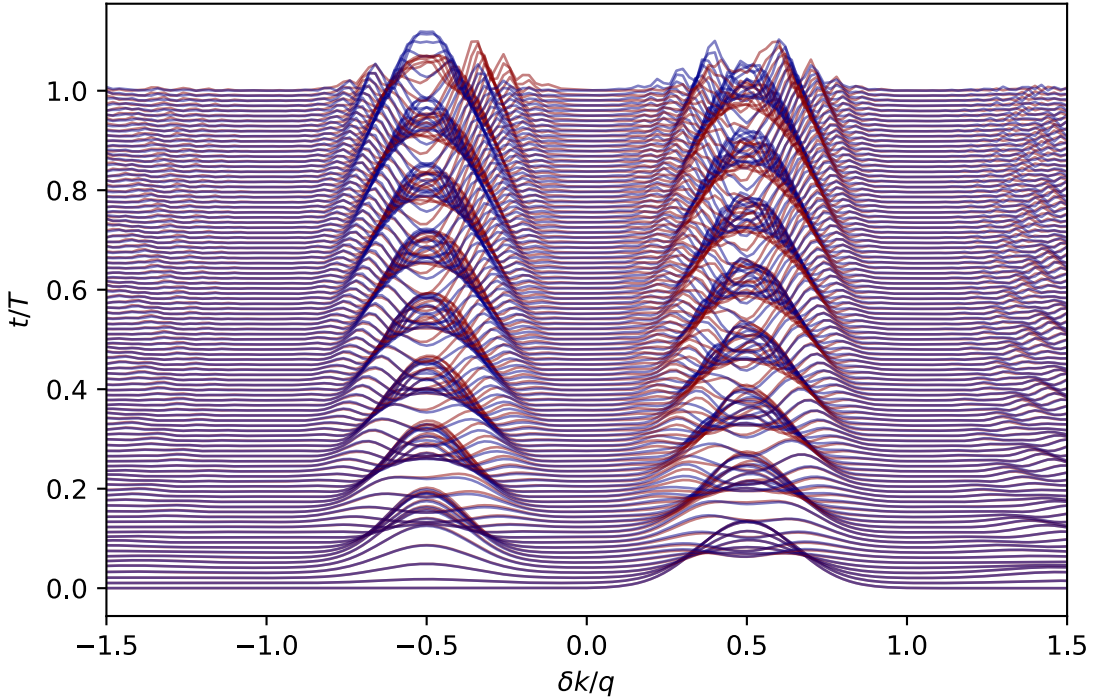

**Fig. S1:** TDSE simulation of slow QEWs ($\beta = 0.02c$, $\Delta_k = 0.15q$) under light-free-electron interaction with gradient-on (red) and gradient-off (blue). Interaction time $T = 25.9$ ps.

Correspondingly, for a faster electron ($\beta = 0.05c$), we simulate its dynamics with a narrower momentum sideband $\Delta_k = 0.05q$ (see Fig. S2). The resulting diffraction pattern already resembles that of the **Raman–Nath regime**, characterized by multiple Rabi oscillations or a quantum random walk, and displays the iconic "light-cone" structure in the time domain.

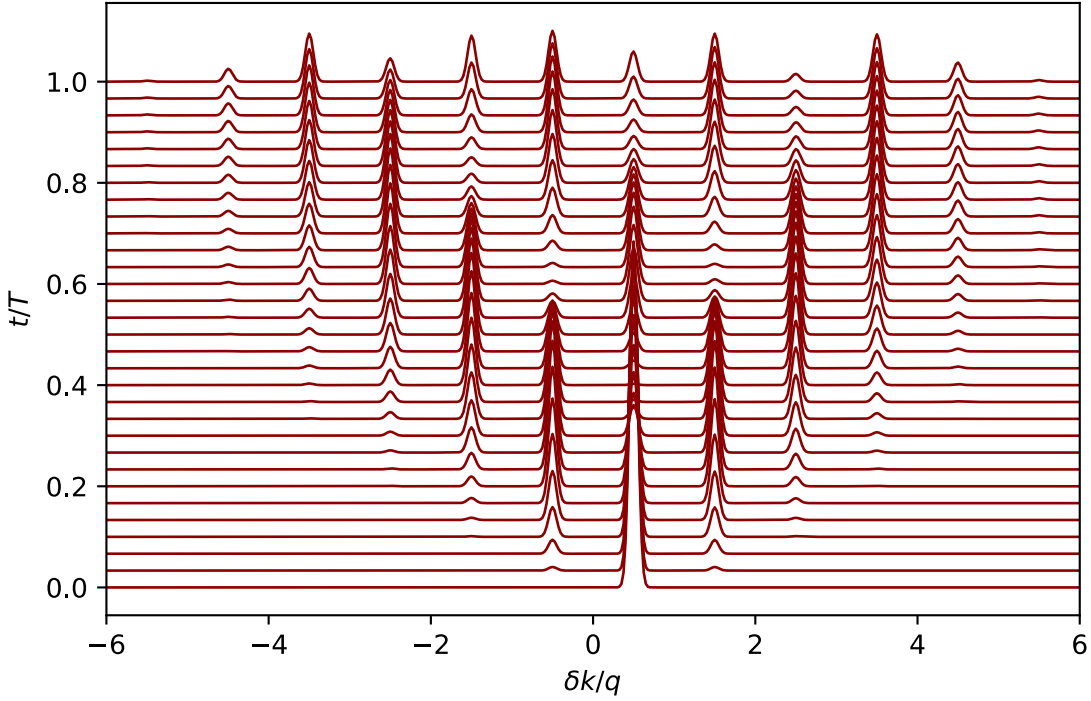

**Fig. S2:** TDSE simulation of a slow QEW ($\beta = 0.05c$, $\Delta_k = 0.05q$) under light-free-electron interaction with the light field $E_0 = 10^8$ V/m. Interaction time $T = 0.52$ ps.

The **Dielectric Laser Accelerator (DLA) regime** is more subtle to realize. In our model, it requires strong spatial localization of the QEW and acceleration by a stronger field. In Fig. S*3*, we initialize the QEW at $\delta k = 0$ and set the phase $\theta = -\pi$ in Eq. (S10), thereby realizing the $\Omega\sigma_y$ term in the effective Dirac Hamiltonian discussed in the main text, instead of the conventional $\Omega\sigma_x$. Other choices of $\theta$ lead to classical bunching ($\theta = \pi/2$) or anti-bunching ($\theta = -\pi/2$), rather than diffraction.

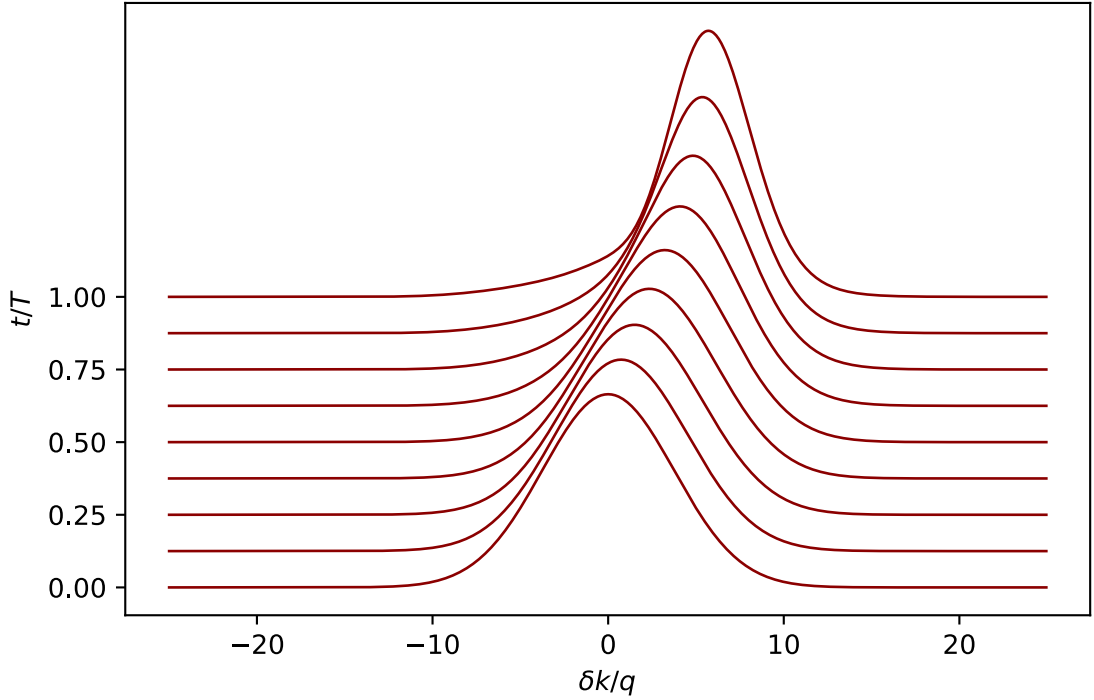

**Fig. S3:** TDSE simulation of a particle-like QEW ($\beta = 0.02c$, $\Delta_k = 1.5q$) accelerated by the stronger light field $E_0 = 10^9$ V/m. Interaction time $T = 0.14$ ps.

Finally, we present the **anomalous PINEM** (APINEM) regime, realized through pre-chirping rather than dispersion. In Fig. S4, we propagate a particle-like QEW with $\Delta_k = 1.2q$, centered at $\delta k = 0$, over a drift distance of $L_D = 10$ cm. This pre-chirping induces an additional phase on the QEW given by $\phi(k) = \hbar k^2 L_D/(2\gamma^3 m)$. The resulting diffraction pattern closely resembles the fishbone-like structure observed by going beyond anomalous Bragg regime with a more broadened momentum sideband. However, in this case, the overlap of sidebands arises from a distinct mechanism, namely, the quadratic phase acquired through free-space propagation.

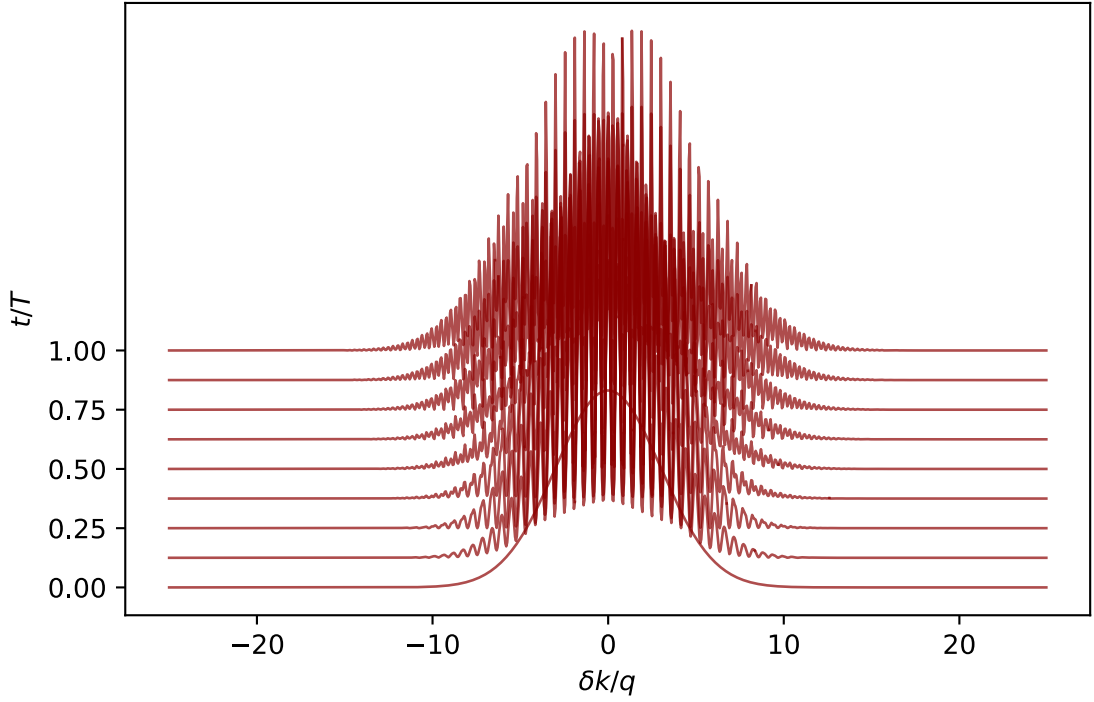

**Fig. S4:** TDSE simulation of a pre-chirped QEW ($\beta = 0.02c$, $\Delta_k = 1.2q$, $L_D = 10$ cm) diffracted by the stronger light field $E_0 = 10^9$ V/m. Interaction time $T = 0.14$ ps.

In this way, we have simulated the dynamics of all diffraction regimes and the corresponding typical patterns mentioned in Table I from the main text.